\documentclass[twocolumn,prl,nobalancelastpage,aps,10pt]{revtex4-1}
\usepackage{graphicx,bm,times}
\newcommand{\squeezeup}{\vspace{-10pt}}

\begin{document}

\title{Atomic Resolution Imaging and Measurement of the Local Density of States of Graphite, Gold and Silicon using Scanning Tunnelling Microscopy}

\author{Ahmed Abdelwahab}

\affiliation{Level 3 MSci. Laboratory, Department of Physics, University of Bristol.}

\begin{abstract} A series of experiments were conducted using the STM instrument, which involves a conducting tip probe to analyse sample surfaces by measurements of a tunnelling current. In this experiment,  STM was used to (1) determine the lattice constant of Highly Oriented Pyrolytic Graphite (HOPG) by acquiring atomic resolution images of its surface, (2) measure the work functions of gold  (Au) and HOPG samples using the STS mode and, (3) compare the variation of the Local Density of States (LDOS) of gold, graphite and Silicon (Si) samples with respect to bias voltage \textit{V}. Experimental values of the lattice constant of HOPG and work functions for gold and graphite were determined  as 0.27 $\pm$ 0.2 nm, 0.7 $\pm$ 0.1 $eV$ and 0.5 $\pm$ 0.1 $eV$ respectively. The lattice constant deviated slightly from the literature value of 0.246 nm, whereas the work functions deviated significantly from the literature values of 5.40 $eV$ for gold and 4.62 $eV$ for graphite. The LDOS for gold was found to be the highest, followed by graphite, then silicon. These findings will be discussed.

\end{abstract}

\date{{December 2019}}

\maketitle

\section{INTRODUCTION}

The Scanning Tunnelling Microscope (STM) is used to obtain atomic resolution images of material surfaces. In 1983, Gerd Binning and Heinrich Rohrer developed the STM at the IBM research laboratory in Switzerland and obtained atomic images of the reconstructed surface of Si(111)-7x7 [1]. STM enables imaging of material surfaces at atomic scales and probing their structure by measuring the variation of the tunnelling current between a conducting tip and a material's surface. As such, STM is a very powerful technique. In this experiment, atomic resolution images of HOPG (Highly Oriented Pyrolytic Graphite) will be obtained, and a lattice constant for HOPG will be determined. The work functions of HOPG and gold (Au) will also be determined. Lastly, graphical analysis will be done to compare the Local Density of States (LDOS) of HOPG, Au and Silicon (Si).

\section{THEORY AND EXPERIMENTAL METHOD}

STM exploits the quantum mechanical phenomenon of electron tunnelling. A conducting tip is brought a few nanometres apart from the sample's surface to be examined. A bias voltage \textit{V} applied between the sample, and the tip allows the electrons to tunnel through the vacuum gap producing a tunnelling current which decays exponentially with distance from the sample. The tunnelling current can be described by the following equation [2]:

\begin{equation}
\textit{I}(z)  = \textit{I}(0) \exp{(-10.2\,\sqrt{\phi_{barrier}}\;z)}. \label{Expansionequation}
\end{equation}

\noindent where z is the distance from the sample to the tip (z = 0 at the sample's surface) and $\phi_{barrier}$ is the apparent barrier height which is defined as the average of the work function of the sample and the tip [3]:

\begin{equation}
\phi_{barrier}  = \frac{\phi_{sample}+\phi_{tip}}{2}. \label{Averageworkfunction}
\end{equation}

For the imaging section of the experiment, atomic resolution images of graphite were obtained. Graphite consists of many stacked graphene sheets bonded by weak Van der Waals forces [4]. Graphene is a two-dimensional sheet of carbon atoms covalently bonded and arranged in periodic hexagonal arrangements [4]. In one hexagon, the separation between alternate carbon atoms is the lattice constant \textit{a}, and the separation between adjacent carbon atoms is the atomic separation \textit{d}. 

The graphene layers are stacked in one of two arrangements, either in an ABAB pattern or ABCABC pattern [4]. The ABAB pattern is illustrated in Fig.1. The white carbon atoms do not have neighbouring atoms underneath, whereas the grey carbon atoms have neighbouring atoms (black) directly underneath them. The electrical conductivity of graphite's surface, therefore, varies slightly with location on the surface [5].

\begin{figure}
\includegraphics*[width=0.96\linewidth,clip]{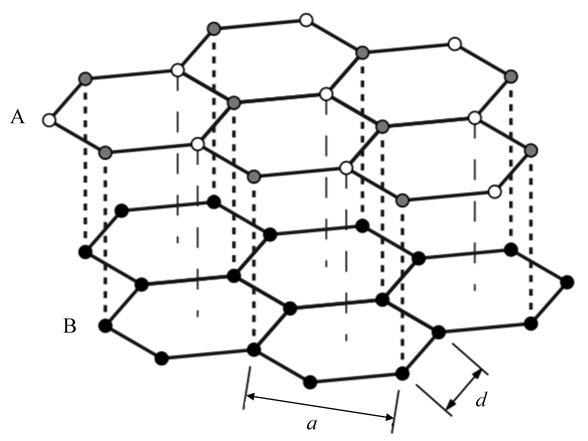}
\caption{ABAB arrangement of the graphene sheets in graphite. The grey carbon atoms have neighbouring atoms directly underneath, whereas the white carbon atoms do not. The lattice constant is \textit{a} and \textit{d} is the atomic separation [5].} \label{expansionfigure}
\end{figure}

The interior angle of a regular hexagon is 120 degrees. So by measuring \textit{d} and following simple trigonometric arguments, the lattice constant \textit{a} can be obtained by the following formula: 

\begin{equation}
\textit{a}  = \sqrt{3}\textit{d}. \label{Latticeconstant}
\end{equation}

Atomic resolution images of the surface of graphite samples were obtained using the constant current mode of STM. In the constant current mode, a PID control system feedback loop maintains the tunnelling current at a fixed setpoint such that the distance \textit{z} is always constant. While maintaining a constant distance, the tip scans the surface in a raster-scan configuration where it first traces a line in the x-axis (fast axis) and then retraces the same line in the opposite direction. After imaging one line, the tip moves up the y-axis (slow axis) and scans another line on the x-axis as before. This process is repeated until an image in the xy plane is obtained showing the variation of the height of the surface's topology as a function of the (\textit{x},\textit{y}) position of the tip on the sample's surface. 

Preparation for the imaging consisted of cutting a sharp gold or platinum-iridium tip (Pt-Ir) and cleaning the graphite sample with scotch tape. The tip and sample are then mounted onto the scan head, and the sample is approached towards the tip until a tunnelling current is detected. A Fourier transform applied on the atomic resolution images allowed for a direct measurement of the atomic separation \textit{d} from which the lattice constant \textit{a} was calculated using Eq.3.

The barrier height $\phi_{barrier}$ is the minimum energy required to remove an electron from the surface of a sample to the vacuum level [2]. $\phi_{barrier}$ depends on the sample's local topology and the states' local density [2]. The spectroscopy mode (STS) was used to determine $\phi_{barrier}$. In STS, a tip (gold) is moved towards the sample's surface, and the tunnelling current is recorded as a function of the distance \textit{z} moved by the tip. An exponential graph, known as an \textit{IZ} curve, modelled by Eq.1, is then obtained. By taking the natural log of both sides of Eq.1, a straight line graph of ln(\textit{I}(\textit{z})) against \textit{z} is produced with a gradient of \textit{A} = $-$10.2$\,\sqrt{\phi_{barrier}}$ which can be rearranged to give an equation for $\phi_{barrier}$:

\begin{equation}
\phi_{barrier}  = {\left(\frac{A}{10.2}\right)}^{2}. \label{phibarrier}
\end{equation}

Several \textit{IZ} curves were recorded, and an average of $\phi_{barrier}$ was obtained for both gold and graphite samples. Determination of $\phi_{gold}$ is straight forward as both the sample and tip are made from gold so, according to Eq.2, $\phi_{barrier}$ = $\phi_{gold}$. In the case of the graphite sample, $\phi_{graphite}$ is obtained by rearrangement of Eq.2:

\begin{equation}
\phi_{graphite}  = 2\,\phi_{barrier}-\phi_{gold}. \label{phigraphite}
\end{equation}

Lastly, the STS mode was used to analyse the LDOS of graphite, gold and silicon samples. Here, the gold tip is kept at a fixed distance \textit{z} from the sample's surface, and the tunnelling current is measured as a function of bias voltage \textit{V}, producing an \textit{IV} curve. The relationship between the tunnelling current and \textit{V} can be written as [6]: 

\begin{equation}
\textit{I}(\textit{V})= -\frac{4\pi e}{\hbar}\rho_{t}(\epsilon)|M(\epsilon)|^{2}\int_{E_{F}-eV}^{E_{F}}d\epsilon \rho_{s}(\epsilon). \label{IVcurrent}
\end{equation}
 
where $|M(\epsilon)|^{2}$ is the matrix element of the tunnelling process, $\rho_{s}(\epsilon)$ and $\rho_{t}(\epsilon)$ are the density of states of the sample and tip respectively, $E_{F}$ is the Fermi level and $\epsilon$ is the state energy. 

A positive bias voltage applied to the tip (while keeping the sample at V = 0) causes the electrons to tunnel from the sample's surface to the tip producing a current \textit{I}(\textit{V}) (with the assumption that there is no tunnelling from the tip to the sample) [6]. Both $\rho_{t}(\epsilon)$ and $|M(\epsilon)|^{2}$ are assumed to be constant with changing energy $\epsilon$ [6]. Differentiating both sides of Eq.6 with respect to V gives the proportionality equation [6]:

\begin{equation}
\frac{d\textit{I}(\textit{V})}{d\textit{V}} \propto \rho_{s}(\epsilon). \label{Conductanceequation}
\end{equation}

Eq.7, therefore, states that the derivative of the \textit{IV} curve is proportional to the local density of states of the sample. The LDOS of the samples will be investigated by comparing their derivatives graphically across a small bias voltage range ($-$V to $+$V).

\section{RESULTS AND DISCUSSION}

Figure 2 shows an atomic resolution image with a superposed regular hexagonal grid. To compare the position of the carbon atoms in the image with graphene's theoretical structure, the hexagons were set with the literature value of \textit{a} = 0.246 nm [7]. There is a good match between the position of the atoms on the grid and the real position of the atoms on the image, but a clear offset appears away from the centre due to image distortion. The distortion can be seen by noticing that the vertical spacing between the atoms is compressed.

\begin{figure}[ht]
\includegraphics*[width=0.70\linewidth,clip]{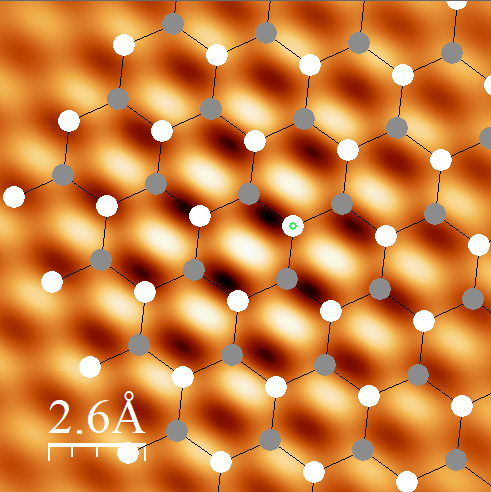}
\caption{Regular hexagon grid fitted on an atomic resolution image of size 1.3 nm (image scale 0.26 nm). The image was self-correlated to show the atomic features with more clarity. The overlap between the grid and the real position of the atoms is offset further away from the centre.} \label{Image}
\end{figure}

Figure 3 shows the Fourier transform diagram of the image. As the diagram is in k-space, measuring the distance between an opposite pair of the brightest pixels yields a value of \textit{d}. As there are three opposite pairs, three values of \textit{d} were obtained, and an average was calculated from which \textit{a} was obtained using Eq.3. Repeating the same analysis on ten images yielded an average experimental value of \textit{a} $=$ 0.27 $\pm$ 0.2 nm.  

\begin{figure}[ht]
\includegraphics*[width=0.70\linewidth,clip]{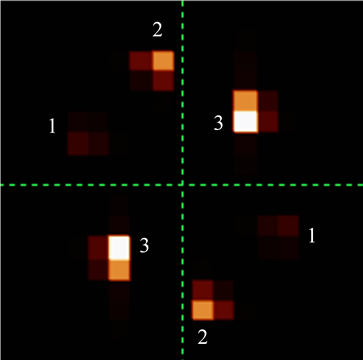}
\caption{Fourier transform diagram of the atomic resolution image in Fig.2. The three opposite pairs are labelled by the numbers 1,2 and 3.} \label{Diagram}
\end{figure}

The experimental value of the lattice constant differs from the literature value of 0.246 nm by 9.3\%. This discrepancy can be attributed to image distortion.

The distortion in the image in Fig.2 is attributed to thermal drift and piezoelectric-electric actuator hysteresis [8]. Thermal drift along the y-axis (slow axis) compressed the vertical spacing between the atoms in the image (the compression is most evident at the top and bottom of the image). The compression is also why the diagram in Fig.3 is compressed in k-space along the diagonal connecting pair 3, which resulted in a high average value for both \textit{d} and \textit{a}. The motion of the tip in the x and y axes is controlled by piezoelectric crystals, which expand or contract by applying a step voltage [8]. Hysteresis occurs when the mechanical displacement of the piezo crystals per voltage step is non-linear, causing non-linear incremental motion of the tip [8]. Hysteresis resulted in stretching the image along the fast scan direction (x-axis), although its effect was not significant as the horizontal spacing between the atoms remained consistent. A paper by Yothers, M.P \textit{et al.} developed a distortion correction method using Matlab to correct thermal drift and hysteresis [8]. The method involves correcting the coordinates of the atoms on an already obtained image using an inverse transform of a computed distortion model, significantly improving the quality of the image [8]. Preparation of sharp tips could also reduce image distortion as tip asymmetry has been shown to distort images [9]. An etching process can produce sharp tips through electrochemical reactions between a tip wire and an electrolyte [2].      

Figure 4 shows a log-log plot for one of the IZ curves from the data for gold and graphite. Using Eq.4, values of the gradient \textit{A} for gold and graphite from this particular log-log plot were 8.84 and 7.82 respectively which yielded values of  $\phi_{barrier}$ of 0.75 \textit{eV} and 0.59 \textit{eV} respectively. For gold, 11 IZ log-log plots were analysed and, since both the sample and tip are made from gold, an average value of $\phi_{barrier}$ = $\phi_{gold}$ = 0.7 $\pm$ 0.1 $eV$ was calculated. \\
\indent As for graphite, an average value of $\phi_{barrier}$ = 0.6 $\pm$ 0.1 \textit{eV} was calculated from 9 IZ log-log plots. By substituting $\phi_{barrier}$ = 0.6 $\pm$ 0.1 \textit{eV} and $\phi_{gold}$ = 0.7 $\pm$ 0.1 $eV$ into Eq.5, an average value of $\phi_{graphite}$ = 0.5 $\pm$ 0.1 \textit{eV} was determined.

\begin{figure}[ht]
\includegraphics*[width=1.0\linewidth,clip]{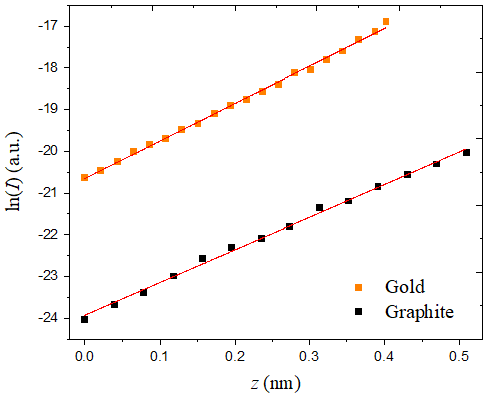}
\caption{ln(\textit{I}) as a function of z for gold and graphite (a.u. arbitrary units). The data was fitted with a straight line of best fit with gradient values 8.84 and 7.82 for gold and graphite, respectively.} \label{IZ}
\end{figure}

The experimental values of the work functions for gold and graphite deviated significantly from the literature values of 5.40 $eV$ for gold and 4.62 $eV$ for graphite [2][10]. The reduction is attributed to the lab temperature and the Pethica mechanism [2]. The quantity $\phi_{barrier}$ is defined as the minimum energy required to remove an electron from a conducting surface to the vacuum level [2]. The definition, therefore, implies that the experiment needs to be carried out near vacuum conditions in order to obtain reasonably high values (several $eV$ or more) of $\phi_{barrier}$ [11]. The temperatures in the experiment ranged from $20^{\circ}$ to $24^{\circ}$ C, suggesting that a lower amount of energy was needed for the electrons to tunnel from the gold and graphite sample surfaces to the tip. This reduced the values of the gradients \textit{A} from the \textit{IZ} log-log plots hence reducing $\phi_{barrier}$, $\phi_{gold}$ and $\phi_{graphite}$. The Pethica mechanism causes more reduction to the value of $\phi_{barrier}$ for the graphite sample. When obtaining \textit{IZ} curves for the graphite sample, the tip touched the sample's surface a few times. As graphite is made of thin graphene sheets, a graphene flake may have attached to the tip. The direct contact between the tip and the flake causes abnormally low values of $\phi_{barrier}$, which further explains the low value for $\phi_{graphite}$ [2].  

Figure 5 shows the LDOS $\rho_{s}(\epsilon)$ of graphite, gold and silicon as a function of bias voltage \textit{V}. This plot is equivalent to plotting $d\textit{I}(V)$$/$$d\textit{V}$ against \textit{V} since $d\textit{I}(V)$$/$$d\textit{V}$ is proportional to the LDOS of the sample $\rho_{s}(\epsilon)$. Since $d\textit{I}(V)$$/$$d\textit{V}$ was obtained by numerical differentiation of the \textit{IV} curves, it was assigned arbitrary units. The graph was plotted across the bias voltage range -30mV to +30mV as the \textit{IV} curves for the samples exhibited a linear behaviour across that range. 

\begin{figure}[ht]
\includegraphics*[width=1.0\linewidth,clip]{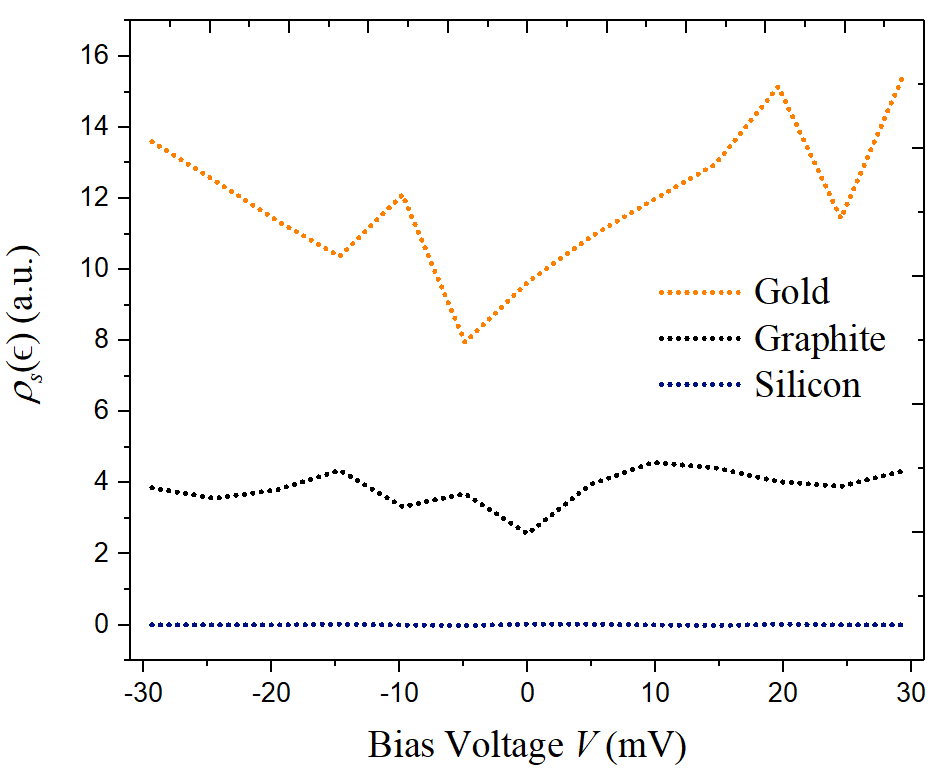}
\caption{$\rho_{s}(\epsilon)$ as a function of \textit{V} for gold (orange), graphite (black) and silicon (dark blue) respectively. The fluctuations in the line profiles are due to noise introduced due to numerically differentiating the \textit{IV} curves data.} \label{IV}
\end{figure}

Figure 5 shows that the LDOS in the restricted bias voltage range -30mV to +30mV is highest for gold, then graphite and finally silicon. As a semiconductor, silicon did not conduct in this restricted range as it has a high band gap of 1.1 $eV$ between its valance band and conduction band, which prevented the electrons from tunnelling from the silicon sample to the tip [12]. As such, the $\rho_{s}(\epsilon)$ values for silicon were approximately zero across this voltage range. The line profiles for gold and graphite were expected to be at a constant $\rho_{s}(\epsilon)$ since their \textit{IV} curves were linear in the restricted bias voltage range. However, numerical differentiation of the \textit{IV} curves of gold and graphite introduced noise in their line profiles. A better data acquisition method is a lock-in amplifier [6]. The lock-in amplifier modulates the bias voltage \textit{V} by a small oscillating voltage \textit{Vsin($\omega$t)} which improves the sensitivity significantly and reduces the noise [6].

\section{CONCLUSIONS}

Analysing atomic resolution images of HOPG, an experimental value for the lattice constant \textit{a} = 0.27 $\pm$ 0.2 was obtained, which is 9.3\% larger than the literature value of 0.246 nm. Log-log analysis of \textit{IZ} curves obtained via the STS mode gave experimental values for the work functions of gold and graphite as $\phi_{gold}$ = 0.7 $\pm$ 0.1 $eV$ and $\phi_{graphite}$ = 0.5 $\pm$ 0.1 $eV$ which deviates significantly from the literature values $\phi_{gold}$ = 5.40 $eV$ and $\phi_{graphite}$ = 4.62 $eV$. A comparison plot of the variation of the LDOS $\rho_{s}(\epsilon)$ with bias voltage \textit{V} showed the highest value of $\rho_{s}(\epsilon)$ was for gold followed by graphite and finally silicon. Silicon did not conduct in the restricted range of the LDOS plot due to its wide band gap of 1.1 $eV$ between its valence band and conductance band, which prevented the electrons from tunnelling from the silicon sample to the tip.    

\section{REFERENCES}
\squeezeup
\squeezeup

\end{document}